\documentclass[aps,prc,twocolumn,showpacs,superscriptaddress]{revtex4}

\usepackage{graphicx}
\usepackage{dcolumn}

\begin{document}


\title{Strong Decays of Light Vector Mesons}



\author{Dennis Jarecke}
\email[]{jarecke@cnr.physics.kent.edu}
\affiliation{Department of Physics, Center for Nuclear Research,
Kent State University, Kent, OH 44240}

\author{Pieter Maris}
\email[]{pmaris@unity.ncsu.edu}
\affiliation{Department of Physics, 
North Carolina State University, Raleigh,  NC 27695-8202}

\author{Peter C. Tandy}
\email[]{tandy@cnr2.kent.edu}
\affiliation{Department of Physics, Center for Nuclear Research,
Kent State University, Kent, OH 44240}

\date{\today}

\begin{abstract}
The vector meson strong decays $\rho\to\pi\pi$, $\phi\to KK$, and
$K^\star \to\pi K$ are studied within a covariant approach based on
the ladder-rainbow truncation of the QCD Dyson--Schwinger equation for
the quark propagator and the Bethe--Salpeter equation for the mesons.
The model preserves the one-loop behavior of QCD in the ultraviolet,
has two infrared parameters, and implements quark confinement and
dynamical chiral symmetry breaking.  The 3-point decay amplitudes are
described in impulse approximation.  The Bethe--Salpeter study
motivates a method for estimating the masses for heavier mesons within
this model without continuing the propagators into the complex plane.
We test the accuracy via the $\rho$, $\phi$ and $K^\star$ masses and 
then produce estimates of the model results for $m_{a_1}$ and 
$m_{b_1}$ as well as the proposed exotic vector $\pi_1(1400)$.

\end{abstract}

%

\maketitle

\section{\label{sec:intro}
Introduction}
The study of light-quark pseudoscalar and vector mesons is an
important tool for understanding how QCD works in the non-perturbative
regime.  The pseudoscalars are important because they are the lightest
observed hadrons and are the Goldstone bosons associated with
dynamical chiral symmetry breaking.  The ground state vector mesons
are important because, as the lowest spin excitations of the pseudoscalars, 
they relate closely to hadronic $\bar{q}q$ modes that are electromagnetically 
excited.
The decay of vector mesons to a pair of pseudoscalars proceeds via a
P-wave interaction; this probes a different aspect of the meson
Bethe--Salpeter [BS] amplitudes than does the electroweak decay
constant which is essentially the projection of the relative
wavefunction onto the origin of separation.

To calculate these coupling constants, we use an approach based on the
Dyson--Schwinger equations [DSEs], which form an excellent tool to
study nonperturbative aspects of hadron properties in QCD
\cite{Roberts:2000aa}.  The approach is consistent with quark and
gluon confinement
\cite{Roberts:2000aa,Burden:1992gd,Krein:1992sf,Maris:1995ns},
generates dynamical chiral symmetry breaking
\cite{Atkinson:1988mw,Roberts:1990mj}, and is Poincar\'e invariant.
It is straightforward to implement the correct one-loop
renormalization group behavior of QCD \cite{Maris:1997tm}, and obtain
agreement with perturbation theory in the perturbative region.
Provided that the relevant Ward--Takahashi identities [WTIs] are
preserved in the truncation of the DSEs, the corresponding currents
are conserved.  Axial current conservation induces the Goldstone
nature of the pions and kaons~\cite{Maris:1998hd}; electromagnetic
current conservation produces the correct electric charge of the
mesons without
fine-tuning.  These properties are implemented here within the rainbow
truncation of the DSE for the dressed quark propagators together with
the ladder approximation for the Bethe--Salpeter equation [BSE] for
meson bound states.  The model we use has two infrared parameters.
Previous work has shown this model to provide an efficient description
of the light-quark pseudoscalar and vector
mesons~\cite{Maris:1997tm,Maris:1999nt}.  Furthermore, in impulse
approximation, the elastic charge form factors of the
pseudoscalars~\cite{Maris:2000sk} and the electroweak transition form
factors of the pseudoscalars and vectors~\cite{Maris:2002mz,Ji:2001pj}
are in excellent agreement with data.  This suggests that the strong
decays of the vector mesons should be well-described in impulse
approximation without parameter adjustment.

The Euclidean metric that we employ facilitates the modeling of the
gluon 2-point function or more generally the effective quark-quark
interaction; it also is the metric within which practical DSE
solutions are available in the literature.  A complicating element is
that solution of the BSE for meson bound states requires an analytic
continuation in the meson momentum to reach the mass-shell.  The
resulting quark $p^2$ in the Bethe--Salpeter integral covers a
certain domain in the complex plane that differs from the real
spacelike axis by an amount that increases with meson mass.  For low
mass mesons such as $\pi$ and $K$, the required continuation of DSE
solutions for the quark propagator is not difficult.  For the ground
state vector mesons, the difficulties are much greater but have been
overcome directly~\cite{Maris:1999nt}.  For meson masses greater than
about $1~{\rm GeV}^2$ the numerical requirements of this procedure
become burdensome.  Furthermore, complex plane singularities of
rainbow DSE solutions for propagators limit this direct approach to
mesons with masses below about 1.2~GeV~\cite{Ahlig:2000qu}.

Here we test the reliability of an approximation that uses a Taylor
expansion to make the continuation away from the real spacelike $p^2$
axis.  Since information about quark propagators is then needed only 
on the
real spacelike $p^2$ axis, this approximation is considerably easier
to implement.  It is similar in spirit to an approximation employed
some time ago~\cite{Jain:1993qh}.  If the behavior for real momenta
dominates the physics, low orders of a Taylor expansion should be
effective.  This work is an exploration of that conjecture since we
can test against the direct complex plane results~\cite{Maris:1999nt}
for BSE solutions for masses of $\rho, K^\star$ and $\phi$ within the
same model.  Independently of how the behavior of the propagators in
the complex plane is obtained, the calculation of meson form factors
or decays in impulse approximation requires that the relative quark
momentum of each vertex amplitude be also continued to the complex
plane.  In the interests of consistency, we employ a Taylor expansion
method for both vertex amplitudes and propagators in the calculation
of decays.

The results from the Taylor expansion method are used to motivate a
method for estimating the masses and BSE solutions for heavier mesons
within this model without continuing the propagators into the complex
plane.  We test the accuracy of this real-axis approximation for the
ground state pseudoscalar and vector mesons, and then produce
estimates of the model results for $m_{a_1}$ and $m_{b_1}$ as well as
the proposed exotic $1^{-+}$ vector $\pi_1(1400)$.

In Sec.~II we review the formulation that underlies a description of
vector meson strong decays within a modeling of QCD through the DSEs
and BSEs.  We discuss the diagrams needed for a calculation in the
impulse approximation for $\rho\to\pi\pi$, $\phi\to KK$, and $K^\star
\to K\pi$.  In Sec.~III we discuss the details of the model and we
include a comparison with recent lattice-QCD results for the quark
propagator.  In Sec.~IV we present our numerical results for masses
and decays, and discuss the performance of the calculational
technique.  In Sec.~V the technique for estimating the masses of heavier
states is evaluated.  A concluding discussion is given in Sec.~VI.

\section{\label{sec:vdecays}
Vector Meson Strong Decays}
We use the Euclidean metric where \mbox{$\{\gamma_\mu,\gamma_\nu\} =
2\delta_{\mu\nu}$}, \mbox{$\gamma_\mu^\dagger = \gamma_\mu$} and
\mbox{$a\cdot b = \sum_{i=1}^4 a_i b_i$}.  The invariant amplitude for
the coupling of a vector state with helicity $\lambda$ and momentum
$p_1 + p_2$ to a pair of pseudoscalars (or scalars) having momenta
$p_{1/2}$ has the form \mbox{${\cal M}^\lambda = \Lambda_\mu \,
\epsilon^\lambda_\mu $} where $\epsilon^\lambda_\mu$ is the
polarization vector of the vector state, and the vertex which
describes the coupling of the Lorentz component of the vector state to
the pair of pseudoscalars has the form \mbox{$\Lambda_\mu = g\, (p_1 -
p_2)_\mu$} where $g$ is the coupling constant.  In the case that the
vector state and the pseudoscalar states are $\bar{q} q$ bound states
with flavor labels $a\bar{b}$, $c\bar{a}$ and $b\bar{c}$ respectively,
the impulse approximation for $\Lambda^{ab,c}_\mu$ can be expressed as
\begin{eqnarray}
\Lambda^{ab,c}_\mu(P;Q) &=& {\rm tr}_{\rm sc} \int^\Lambda_k \! 
        S^c(q) \, \Gamma_{ps}^{c\bar{a}}(q,q_+) \, S^a(q_+)  
\nonumber \\ && 
        \times \Gamma^{a\bar{b}}_\mu(q_+,q_-)\, S^b(q_-) \,
        \bar\Gamma^{b\bar{c}}_{ps}(q_-,q) \;,
\label{triangle}
\end{eqnarray}
where \mbox{$q = k+P/2$} and \mbox{$q_\pm = k-P/2 \pm Q/2$}, and 
${\rm tr}_{sc}$ denotes a trace over color and Dirac spin.  For
convenience we have introduced \mbox{$Q=p_1 + p_2$} as the momentum 
of the vector, and \mbox{$P=(p_1 - p_2)/2$}.  We denote the properly
normalized BS amplitude at a vertex having an outgoing quark of flavor
$f_1$ and momentum $p_+$ and an incoming quark of flavor $f_2$ and
momentum $p_-$ by $\Gamma^{f_1 \bar{f}_2}(p_+,p_-)$.  The notation
\mbox{$\int^\Lambda_k \equiv \int^\Lambda d^4 k/(2\pi)^4$} stands for
a translationally invariant regularization of the integral, with
$\Lambda$ being the regularization mass-scale.  The regularization can
be removed at the end of all calculations, by taking the limit
\mbox{$\Lambda \to \infty$}.  It is understood that the same
regularization is applied to the calculated dressed quark propagators
$S(q)$ and BS amplitudes appearing in Eq.~(\ref{triangle}), and that
the quark propagators are renormalized at a convenient spacelike
momentum scale before taking the limit \mbox{$\Lambda \to \infty$}.

For mesons with a unique quark flavor content, the physical vertices
are given by the $\Lambda^{ab,c}_\mu(P;Q)$.  For states like
\mbox{$|\rho^0>=(|u\bar{u}>-|d\bar{d}>)/\sqrt{2}$}, linear
combinations are required.  We have, for example,
\begin{eqnarray}
\label{vertex_rpp}
\Lambda^{\rho^0 \to \pi^+\pi^-}_\mu &=& \frac{1}{\sqrt{2}} 
         [\Lambda^{uu,d}_\mu(P;Q)-\Lambda^{dd,u}(-P;Q)],\\
\label{vertex_pkk}
\Lambda^{\phi \to K^+K^-}_\mu &=& \Lambda^{ss,u}_\mu(P;Q),\\ 
\label{vertex_kstar+1}
\Lambda^{K^{\star+} \to K^+\pi^0}_\mu &=& \frac{1}{\sqrt{2}}
                                \Lambda^{us,u}_\mu(P;Q),\\
\label{vertex_kstar+2}
\Lambda^{K^{\star+} \to K^0\pi^+}_\mu &=& \Lambda^{us,d}_\mu(P;Q)~.
\end{eqnarray}
The vertices for modes that differ from these by electric charge are
obtained by quark flavor exchange.  For example under isospin exchange
of $u$ and $d$ flavors, the last two modes become \mbox{$K^{\star0}
\to K^0 \pi^0$} and \mbox{$K^{\star0} \to K^+ \pi^-$} respectively.
In this work we use isospin symmetry in which $u$- and $d$-quarks have
identical strong interactions and properties except electric charge.
Under these circumstances, \mbox{$\Lambda^{us,u}_\mu =$}
\mbox{$\Lambda^{ds,d}_\mu =$} \mbox{$\Lambda^{us,d}_\mu$} and there is
only one independent $K^\star \to K \pi$ vertex.
 
The physical coupling constants for each \mbox{$v \to p p$} decay are
identified from the vertices at the mass shell according to
\begin{equation}
\Lambda^{v \to p p}_\mu(P;Q^2=-m_v^2) 
        \; = \; 2 P_\mu^T \, g_{v \to p p} ~, 
\label{g_rpp}
\end{equation}
where $P^T$ is the component of $P$ perpendicular to the vector meson
momentum $Q$.  The decay width of a vector of mass $m_v$ is given by
\begin{equation}
\label{genwidth}
\Gamma \; = \; \frac{1}{3} \sum_{\lambda=1}^3 \! 
                    \frac{\hat{\rho}}{2\,m_v} |{\cal M}^\lambda|^2 ~,
\end{equation}
where $\hat{\rho}$ is the invariant phase space factor, the invariant 
amplitude is \mbox{${\cal M}^\lambda = \Lambda_\mu \, \epsilon_\mu^\lambda$},
and $\epsilon_\mu^\lambda$ is the polarization vector.  The explicit form is
\begin{eqnarray}
\label{width}
\Gamma \; = \; \frac{g^2}{48\pi\, m_v^5}\; 
        \lambda^{3/2}(m_v^2,m^2_{p1},m^2_{p2})~,
\end{eqnarray}
where
\begin{equation}
\lambda(a,b,c) \; = \; a^2 + b^2 + c^2 -2ab -2ac -2bc~,
\label{lam}
\end{equation}
and $m_{p1}$, and $m_{p2}$ are the masses of the pseudoscalar decay
products.

\subsection{\label{sec:DSEs} Dyson--Schwinger Equations}
The dressed quark propagator and the meson BS amplitudes are solutions
of their respective DSEs, namely
\begin{eqnarray}
\lefteqn{S(p)^{-1}\;=\;Z_2 \, i\,/\!\!\!p + Z_4 \, m(\mu)}
\nonumber\\ &&
        {} + Z_1 \int^\Lambda_q \! g^2D_{\mu\nu}(p-q) \, 
        \frac{\lambda^i}{2}\gamma_\mu \, S(q) \, \Gamma^i_\nu(q,p)~,
\label{quarkdse}
\end{eqnarray}
and
\begin{eqnarray}
\Gamma^{a\bar{b}}(p_+,p_-) &=& \int^\Lambda_q \! K(p,q;Q)S^a(q_+)
                                          \Gamma^{a\bar{b}}(q_+,q_-)S^b(q_-),
\nonumber\\ {}
\label{bse}
\end{eqnarray}
where $D_{\mu\nu}(k)$ is the renormalized dressed-gluon propagator,
$\Gamma^i_\nu(q,p)$ is the renormalized dressed quark-gluon vertex,
and $K$ is the renormalized $q\bar{q}$ scattering kernel that is
irreducible with respect to a pair of $q\bar{q}$ lines. The quark
momenta are $q_\pm$; the meson momentum is \mbox{$P= q_+ - q_-$} and
satisfies $P^2 = -m^2$.

The solution of Eq.~(\ref{quarkdse}) is renormalized according to
$S(p)^{-1}=i\gamma\cdot p+m(\mu)$ at a sufficiently large spacelike
$\mu^2$, with $m(\mu)$ the renormalized quark mass at the scale $\mu$.
In Eq.~(\ref{bse}), $S$, $\Gamma^i_\mu$, and $m(\mu)$ depend on the
quark flavor, although we have not indicated this explicitly.  The
renormalization constants $Z_2$ and $Z_4$ depend on the
renormalization point and the regularization mass-scale, but not on
flavor: in our analysis we employ a flavor-independent renormalization
scheme.

To implement Eq.~(\ref{bse}) while preserving the constraint \mbox{$P=
q_+ - q_-$}, it is necessary to specify how the total momentum is
partitioned between the quark and the antiquark.  The general choice
is \mbox{$q_+ =$} \mbox{$q+\eta P$} and \mbox{$q_- =$} \mbox{$q -$}
\mbox{$(1-\eta) P$} where $\eta$ is the momentum partitioning
parameter.  The choice of $\eta$ is equivalent to a choice of relative
momentum $q$; physical observables should not depend on the choice.
This provides us with a convenient check on numerical methods.
 
The meson BS amplitude $\Gamma^{a\bar{b}}$ is normalized according to
the canonical normalization condition
\begin{eqnarray}
2\, P_\mu &=& {\rm tr}_{\rm sc} \, \frac{\partial}{\partial P_\mu} 
        \Big\{ 
  \int^\Lambda_q \! \bar\Gamma^{a\bar{b}}(\tilde{q}',\tilde{q})\,S^a(q_+)\,
        \Gamma^{a\bar{b}}(\tilde{q},\tilde{q}')\,S^b(q_-) 
\nonumber\\ &&
        {} + \int^\Lambda_{k,q} \!
        \bar\chi^{a\bar{b}}(\tilde{k}',\tilde{k})\,K(k,q;P)\,
        \chi^{a\bar{b}}(\tilde{q},\tilde{q}') \Big\} \, ,
\label{gennorm}
\end{eqnarray}
at the mass shell ${P^2=Q^2=-m^2}$, with \mbox{$\tilde{q}=q+$}
\mbox{$\eta Q$}, \mbox{$\tilde{q}'=q-(1-\eta) Q$}, and similarly for
$\tilde{k}$ and $\tilde{k}'$.  Thus the derivative in
Eq.~(\ref{gennorm}) acts only upon the propagators $S$ and the kernel
$K$.  For vector mesons, it is understood that one must contract and
average over the Lorentz indices of the (transverse) BS amplitudes due
to the three independent polarizations.

For pseudoscalar bound states the BS amplitude is commonly decomposed
into~\cite{Maris:1997tm}
\begin{eqnarray}
\label{genpion}
\Gamma(q_+, q_-) = 
        \gamma_5 \big[ i E + \;/\!\!\!\! P \, F + \,/\!\!\!q \, G +
        \sigma_{\mu\nu}\,q_\mu P_\nu \,H \big]\,,
\end{eqnarray}
where the 4 independent covariants have been constructed from the
momentum basis $q$ and $P$ rather than from $q_+$ and $q_-$.  Hence
the invariant amplitudes $E$, $F$, $G$ and $H$ are Lorentz scalar
functions $f(q^2,q\cdot P;\eta)$ that have an explicit dependence upon
the momentum partitioning parameter $\eta$.  The dependence of the
amplitudes upon \mbox{$q\cdot P$} can be conveniently represented by
the following expansion based on Chebyshev polynomials
\begin{equation}
\label{expansion}
\label{chebmom}
 f(q^2, q\cdot P) \; = \; \sum_{i=0}^\infty
        U_i(\cos\theta) (q P)^i f_i(q^2) \,,
\end{equation}
where \mbox{$\cos\theta = q\cdot P /(q P)$}.  For charge-parity
eigenstates such as the pion, each amplitude $E$, $F$, $G$, and $H$
will have a well-defined parity in the variable $q\cdot P$ if one
chooses $\eta=1/2$.  In this case, these amplitudes are either
entirely even ($E$, $F$, and $H$) or odd ($G$) in $q\cdot P$, and only
the even ($E$, $F$, and $H$) or odd ($G$) Chebyshev moments $f_i$ are
needed for a complete description.

Since a massive vector meson bound state is transverse, the BS
amplitude requires eight covariants for its representation.  We choose
the transverse projection of the form
\begin{eqnarray}
\lefteqn{ \Gamma_\mu(q_+,q_-) \;=\;
        \gamma_\mu \, V_1 + q_\mu \,/\!\!\!q \, V_2
        + q_\mu \;/\!\!\!\! P \, V_3 }
\nonumber \\   && {} 
        + \gamma_5\epsilon_{\mu\alpha\nu\beta}\gamma_\alpha 
                q_\nu P_\beta \, V_4
        + q_\mu \, V_5
        + \sigma_{\mu\nu}q_\nu  \, V_6
\nonumber \\   && {} 
        + \sigma_{\mu\nu}P_\nu \, V_7
        + q_\mu \sigma_{\alpha\beta}q_\alpha P_\beta \, V_8 \;.
\label{vecBSAform}
\end{eqnarray}
This form is a variation of that used in Ref.~\cite{Maris:1999nt} that
is simpler and easier to use in many respects.  The invariant
amplitudes $V_i$ are Lorentz scalar functions of $q^2$ and $q\cdot P$
and, for charge eigenstates, they are either odd or even in $q\cdot
P$.  For the $1^{--}$ rho meson, $V_3$ and $V_6$ are odd, the other
amplitudes are even.  To reverse the charge parity, one simply
reverses this odd-even property of the amplitudes.

\section{\label{sec:ladrain}
Ladder-rainbow model}
We employ the model that has been developed recently for an efficient
description of the masses and decay constants of the light
pseudoscalar and vector mesons~\cite{Maris:1997tm,Maris:1999nt}.  This
consists of the rainbow truncation of the DSE for the quark propagator
and the ladder truncation of the BSE for the pion and kaon amplitudes.
The required effective $\bar q q$ interaction is constrained by
perturbative QCD in the ultraviolet and has a phenomenological
infrared behavior.  In particular, the rainbow truncation of the quark
DSE, Eq.~(\ref{quarkdse}), is
\begin{equation}
\label{ourDSEansatz}
Z_1 g^2 D_{\mu \nu}(k) \Gamma^i_\nu(q,p) \rightarrow
 {\cal G}(k^2) D_{\mu\nu}^{\rm free}(k)\, \gamma_\nu
                                        \textstyle\frac{\lambda^i}{2} \,,
\end{equation}
where $D_{\mu\nu}^{\rm free}(k=p-q)$ is the free gluon propagator in
Landau gauge.  The consistent ladder truncation of the BSE,
Eq.~(\ref{bse}), is
\begin{equation}
\label{ourBSEansatz}
        K(p_+,q_+;P) \to
        -{\cal G}(k^2)\, D_{\mu\nu}^{\rm free}(k)
        \textstyle{\frac{\lambda^i}{2}}\gamma_\mu \otimes
        \textstyle{\frac{\lambda^i}{2}}\gamma_\nu \,,
\end{equation}
where \mbox{$k=p-q$}.  These two truncations are consistent in the sense
that the combination produces vector and axial-vector vertices
satisfying the respective WTIs.  In the axial case, this ensures that in
the chiral limit the ground state pseudoscalar mesons are the massless
Goldstone bosons associated with chiral symmetry
breaking~\cite{Maris:1997tm,Maris:1998hd}.  In the vector case, this ensures
electromagnetic current conservation.

\begin{table}
\caption{\label{tab:sumres}
The pseudoscalar observables that define the present ladder-rainbow DSE-BSE
model, adapted from Refs.~\protect\cite{Maris:1997tm,Maris:1999nt}.  
}
\begin{ruledtabular}
\begin{tabular}{l|cc}
  & \multicolumn{1}{r}{experiment~\protect\cite{Groom:2000in}}
  & \multicolumn{1}{r}{calculated}  \\
  & \multicolumn{1}{r}{(estimates)}
  & \multicolumn{1}{r}{($^\dagger$ fitted)} \\ \hline
$m^{u=d}_{\mu=1 {\rm GeV}}$ &
   \multicolumn{1}{r}{ 5 - 10 MeV}  & \multicolumn{1}{r}{ 5.5 MeV}     \\
$m^{s}_{\mu=1 {\rm GeV}}$ &
   \multicolumn{1}{r}{ 100 - 300 MeV} &\multicolumn{1}{r}{ 125 MeV} \\ \hline
- $\langle \bar q q \rangle^0_{\mu}$
                & (0.236 GeV)$^3$ & (0.241$^\dagger$)$^3$ \\
$m_\pi$         &  0.1385 GeV &   0.138$^\dagger$ \\
$f_\pi$         &  0.131 GeV &   0.131$^\dagger$ \\
$m_K$           &  0.496 GeV  &   0.497$^\dagger$ \\
$f_K$           &  0.160 GeV  &   0.155     \\ 
\end{tabular}
\end{ruledtabular}
\end{table}

The model is completely specified once a form is chosen for the
``effective coupling'' ${\cal G}(k^2)$.  The ultraviolet behavior is
chosen to be that of the QCD running coupling $\alpha(k^2)$; the
ladder-rainbow truncation then generates the correct perturbative QCD
structure of the DSE-BSE system of equations.  The phenomenological
infrared form of ${\cal G}(k^2)$ is chosen so that the DSE kernel
contains sufficient infrared enhancement to produce an empirically
acceptable amount of dynamical chiral symmetry breaking as represented
by the chiral condensate~\cite{Hawes:1998cw}.

We employ the Ansatz found to be successful in earlier 
work~\cite{Maris:1997tm,Maris:1999nt}
\begin{eqnarray}
\label{gvk2}
\frac{{\cal G}(k^2)}{k^2} &=&
        \frac{4\pi^2\, D \,k^2}{\omega^6} \, {\rm e}^{-k^2/\omega^2}
\nonumber \\  &+&
 \frac{ 4\pi^2\, \gamma_m \; {\cal F}(k^2)}
        {\frac{1}{2} \ln\left[\tau +
        \left(1 + k^2/\Lambda_{\rm QCD}^2\right)^2\right]} \,,
\end{eqnarray}
{\sloppy 
with \mbox{$\gamma_m=\frac{12}{33-2N_f}$} and
\mbox{${\cal F}(s)=$} \mbox{$(1 - \exp(\frac{-s}{4 m_t^2}))/s$}.
The first term implements the strong infrared enhancement in the region
\mbox{$0 < k^2 < 1\,{\rm GeV}^2$} required for sufficient dynamical
chiral symmetry breaking.  The second term serves to preserve the
one-loop renormalization group behavior of QCD.  We use
\mbox{$m_t=0.5\,{\rm GeV}$}, \mbox{$\tau={\rm e}^2-1$}, \mbox{$N_f=4$},
and we take \mbox{$\Lambda_{\rm QCD} = 0.234\,{\rm GeV}$}.  The
renormalization scale is chosen to be \mbox{$\mu=19\,{\rm GeV}$} which
is well into the domain where one-loop perturbative behavior is
appropriate~\cite{Maris:1997tm,Maris:1999nt}.  The remaining parameters,
\mbox{$\omega =$} \mbox{$0.4\,{\rm GeV}$} and \mbox{$D=0.93$} 
\mbox{${\rm GeV}^2$} along with the quark masses, are fitted to give 
a good description of $\langle\bar q q\rangle$, $m_{\pi/K}$ and
$f_{\pi}$.  The subsequent values for $f_K$ and the masses and decay
constants of the vector mesons $\rho, \phi, K^\star$ are found to be
within 10\% of the experimental data~\cite{Maris:1999nt}, see
Tables~\ref{tab:sumres} and \ref{tab:resvecnew}.  A detailed analysis
of the relationship between QCD and this Landau gauge, rainbow-ladder
truncation of the DSEs with renormalization group improvement, can be
found in the originating work~\cite{Maris:1997tm}.  }

\begin{figure}[htb]
\hspace*{-0mm}\includegraphics[width=8.0cm]{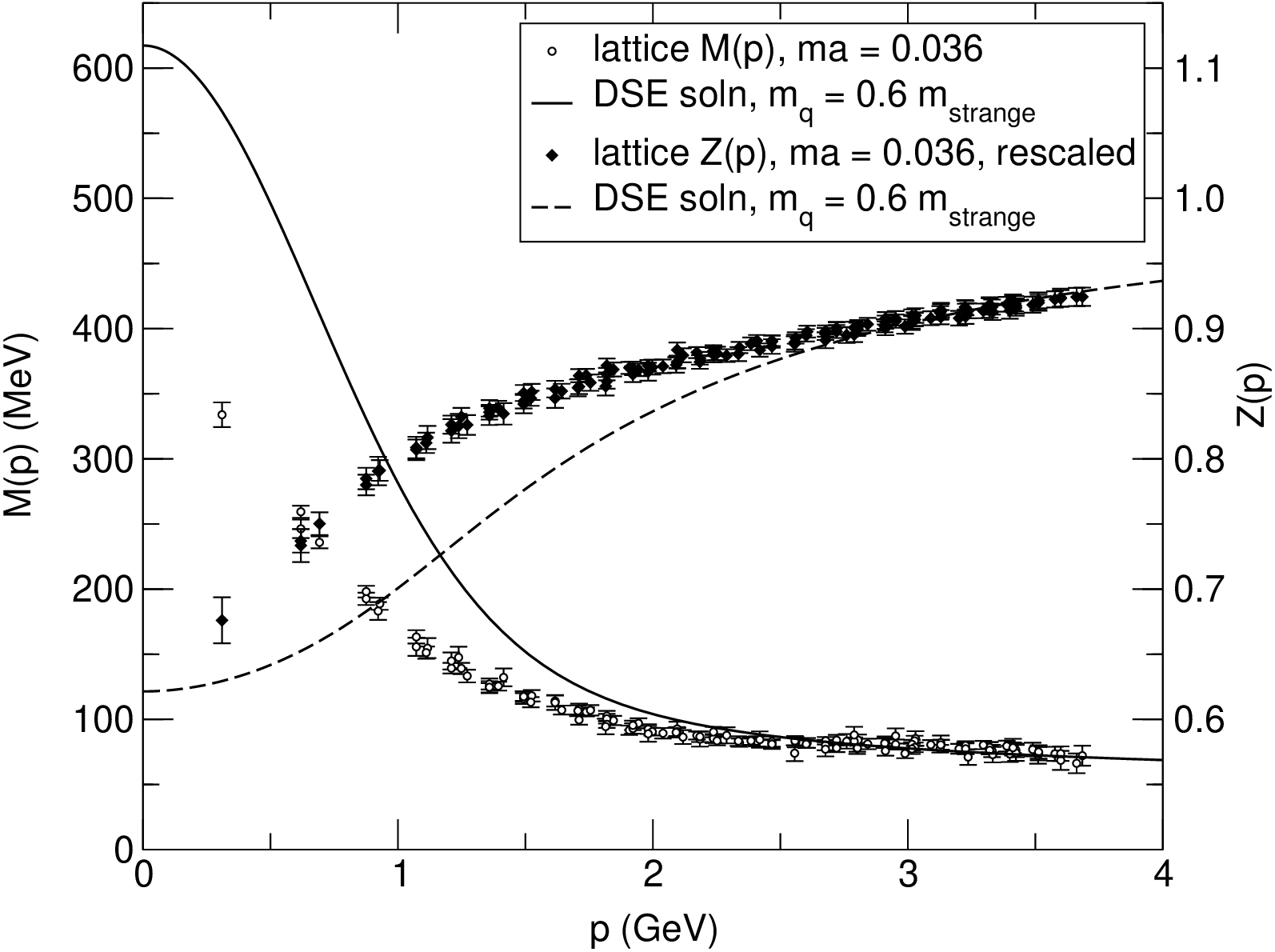}
\caption{DSE solution~\protect\cite{Maris:1999nt} for quark propagator 
amplitudes compared to recent lattice
data~\protect\cite{Bowman:2002bm,Bowman_privCom02}.
\label{fig:MplusZ} }
\end{figure}

\begin{figure}[htb]
\includegraphics[width=8.0cm,angle=0]{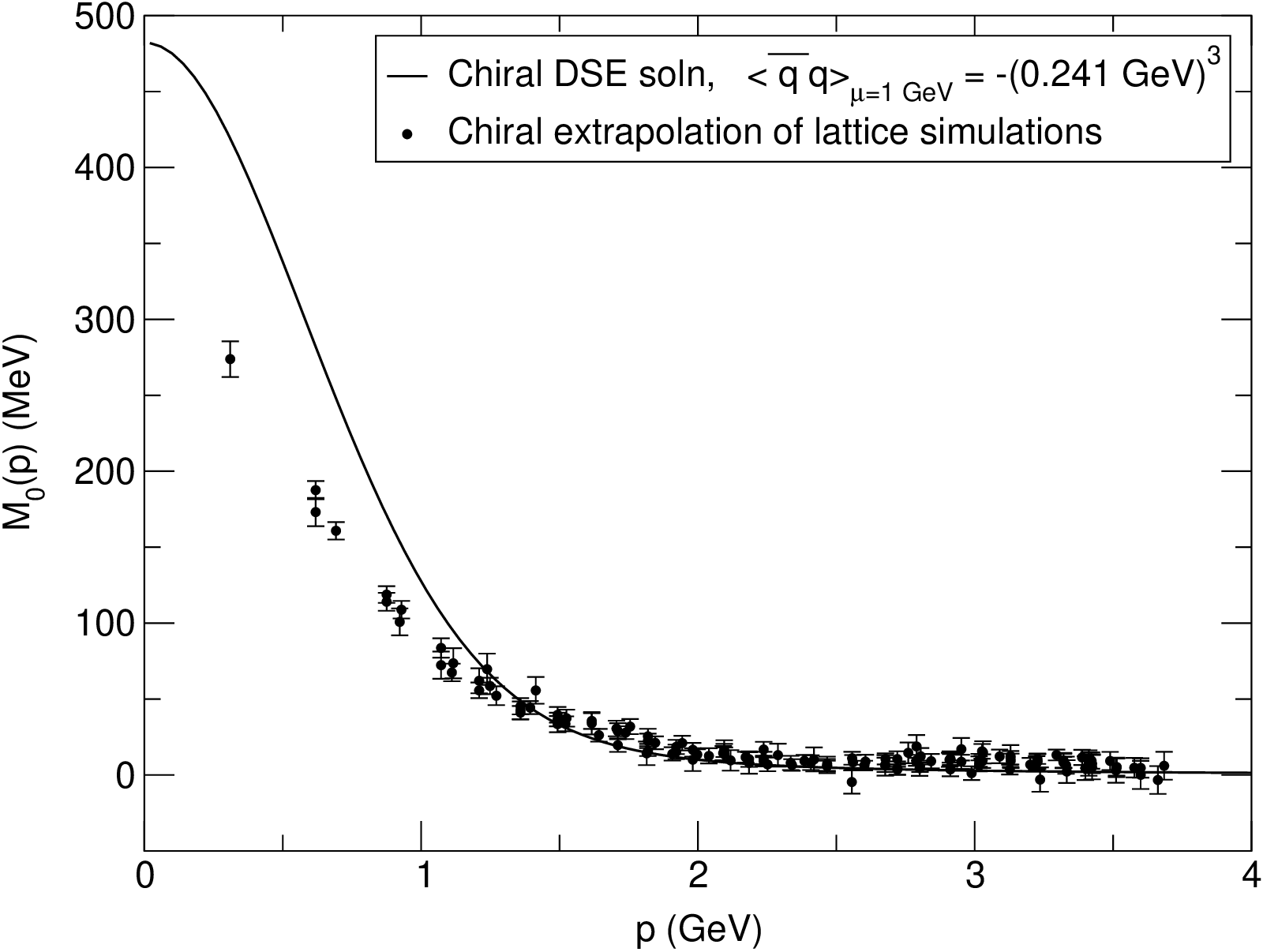}
\caption{The chiral limit DSE mass function compared to the lattice chiral
extrapolation~\protect\cite{Bowman:2002bm,Bowman_privCom02}
\label{fig:Mchiral_Jul02} }
\end{figure}

In Figs.~\ref{fig:MplusZ} and \ref{fig:Mchiral_Jul02} we compare the
DSE model~\cite{Maris:1999nt} propagator amplitudes defined by
\mbox{$S(p) =$} \mbox{$Z(p^2)[ i /\!\!\! p + M(p^2)]^{-1}$} with the
most recent results in lattice QCD using staggered fermions in Landau
gauge~\cite{Bowman:2002bm,Bowman_privCom02}.  These simulations were
done with the Asqtad improved staggered quark action, which has
lattice errors of order ${\cal O}(a^4)$ and ${\cal O}(a^2\,g^2)$.  In
previous such comparisons~\cite{Maris:2000zf,Tandy:2001qk} the lattice
data was less reliable.  Fig.~\ref{fig:MplusZ} shows both $M(p)$ and
$Z(p)$ obtained with a bare lattice mass of $ma = 0.036$ in lattice
units, which corresponds to a bare mass of $57$~MeV in physical units.
The DSE calculations use a current mass value of $75$~MeV at $\mu$ =
$1$~GeV to match the lattice mass function around $3$~GeV; this
current mass is about $0.6 \, m_s$.  There is agreement in the
qualitative infrared structure of the mass function particularly in
the way the infrared enhancement sets in.  Since the lattice
simulation produces the regulated but un-renormalized propagator, the
scale of the field renormalization function $Z$ is arbitrary and only
the shape is a meaningful comparison.  For this reason, we have
rescaled the lattice data for $Z$ so that they match the DSE solution
above $3$~GeV.  The ladder-rainbow DSE model typically produces a $Z$
that saturates much slower than does the lattice $Z$; this may signal
a deficiency of the bare gluon-quark vertex.

The lattice work~\cite{Bowman:2002bm,Bowman_privCom02} also produced a
linear extrapolation to the chiral limit mass function $M_0(p)$ and we
compare this to the DSE result in Fig~\ref{fig:Mchiral_Jul02}.  In
principle the leading UV behavior of this mass function is
proportional to the chiral condensate.  The extraction of this from
earlier lattice data has been attempted but there are uncertainties
still to be resolved~\cite{Bonnet:2002ih}.  We note here that above
$1$~GeV there is excellent agreement between the present DSE model and
the lattice results.

Recent reviews~\cite{Roberts:2000aa,Alkofer:2000wg} put this model in
a wider perspective.  These reviews include a compilation of results
for both meson and baryon physics with similar models, an analysis how
quark confinement is manifest in solutions of the DSEs, and both
finite temperature and finite density extensions.  The question of the
accuracy of the ladder-rainbow truncation has also received some
attention; it was found to be particularly suitable for the flavor
octet pseudoscalar mesons since the next-order contributions in a
quark-gluon skeleton graph expansion, have a significant amount of
cancellation between repulsive and attractive
corrections~\cite{Bender:1996bb}.

\section{\label{sec:results} Results}

\subsection{Method of Calculation}
In order to identify the meson mass $m$ from the BSE, Eq.~(\ref{bse}),
in Euclidean metric, it is convenient to first introduce the linear
eigenvalue $\lambda(P^2)$ of the kernel, continue the equation in
$P^2$ to the timelike region, and then find $m$ such that
\mbox{$\lambda(-m^2) = 1$}.  The BSE kernel involves terms linear in
the 4-vector $P$ and they are given the continuation \mbox{$ P \to
(i\, m, \vec{0})$}.  Thus with the quark propagators written as
\mbox{$S(p) =- i /\!\!\! p \sigma_V(p^2) +$} \mbox{$\sigma_S(p^2)$},
the kernel of the BSE involves the amplitudes
$\sigma_\alpha(q_\pm^2)$, where \mbox{$\alpha=V, S$} and
\mbox{$q_\pm^2 = q^2$} \mbox{$ - m^2/4 \pm i\, q\, m\, z$}.  Here
\mbox{$q = \sqrt{q^2} >0$} with $q^2$ being the (spacelike)
integration variable, \mbox{$-1 < z < 1$} is a direction cosine, and
we have adopted equal momentum partitioning \mbox{$\eta =1/2$} for
convenience.  The amplitudes $\sigma_\alpha(q_\pm^2)$ are in principle
required to be known in a parabolic domain of the complex $q_\pm^2$
plane that includes the positive real axis and that extends
symmetrically in the imaginary direction and in the negative
(timelike) direction by amounts that grow with $m$.  Previous work
within the present approach and model has proceeded by use of the
quark DSE to make the required analytic continuation of these
propagator amplitudes.  For \mbox{$m > 1~{\rm GeV}$}, the difficulty
of the necessary numerical methods can outweigh the benefits of
solution, and certainly accuracy becomes problematic.  For this
reason, we explore an approximate method.

The eigenvalue $\lambda(P^2)$ should be real to obtain a real mass (no
meson decay mechanisms are present in the ladder BSE kernel) and thus
one can speculate that the sub-domain consisting of the positive real
axis for $q_\pm^2$ might dominate the physics.  To explore this 
systematically, we
use a Taylor expansion of $\sigma_\alpha(q_\pm^2)$ to reach the
argument values off the positive real axis.  Thus the only information
needed about the quark propagator amplitudes from solution of the DSE
are their values and derivatives on the Euclidean momentum domain.
Within the BSE kernel we thus employ
\begin{eqnarray}
\label{taylor}
\sigma_\alpha(q_\pm^2)    &=& 
  \sigma_\alpha(q_E^2) + \Delta_\pm^2 \,\sigma_\alpha^\prime(q_E^2)  
        + \cdots \;\;\;\; ,      
\end{eqnarray}
where \mbox{$q_E^2 = {\rm max}[0, \Re(q_\pm^2)]$} and 
\mbox{$q_E^2 + $} \mbox{$\Delta_\pm^2 = q_\pm^2$}.  For practical
reasons we are interested only in low orders of this expansion.

\subsection{Masses and Decay Constants}
\label{masses}
\begin{table}
\caption{\label{tab:PSresults}  Masses and decay constants (in GeV) 
for $\pi$ and $K$ obtained from the Taylor expansion treatment of
$q_\pm^2$ in quark propagator amplitudes.  The effect of truncation  
of the 4 Dirac covariants to just the canonical $\gamma_5$ covariant
is indicated.  Comparison is made with results from use of quark 
amplitudes along $\Re(q_\pm^2)$ only, and with the model-exact 
results.}
\begin{ruledtabular}
\begin{tabular}{l|cccc}
                        & $m_\pi$ & $f_\pi$  & $m_K$ & $f_K$   \\ \hline 
$1^{st}$ order Taylor   &       &          &        &       \\ 
All 4 ampls             &0.117  &0.111     &0.420   &0.134     \\
$E$ ampl only           &0.106  &0.086     &0.383   &0.103     \\ \hline
$2^{nd}$ order Taylor   &       &          &        &         \\ 
All 4 ampls             &0.138  &0.131     &0.498   &0.157    \\
$E$ ampl only           &0.121  &0.098     &0.436   &0.116     \\ \hline
$3^{rd}$ order Taylor   &       &          &        &         \\ 
All 4 ampls             &0.138  &0.131     &0.496   &0.153         \\
$E$ ampl only           &0.121  &0.098     &0.435   &0.114     \\ \hline
Real axis only          &0.123  &0.099     &0.440   &0.127       \\ \hline
Model exact~\protect\cite{Maris:1999nt}  
                        &0.138  &0.131     & 0.497  & 0.155   \\ \hline
Experiment              &0.1385 &0.131     &0.496   &0.160    \\ 
\end{tabular}
\end{ruledtabular}
\end{table}
In Table~\ref{tab:PSresults}, we show the results for the masses and
decay constants of $\pi$ and $K$ through third order in this expansion
compared to the model exact results obtained earlier via direct
analytic continuation of the DSE solutions for propagators.  Agreement
to three significant figures is obtained at second order for $\pi$
while there is evidently still an error of about 2\% in $f_K$ after
third order for $K$.

\begin{figure}[h]
\includegraphics[width=7.5cm,angle=0]{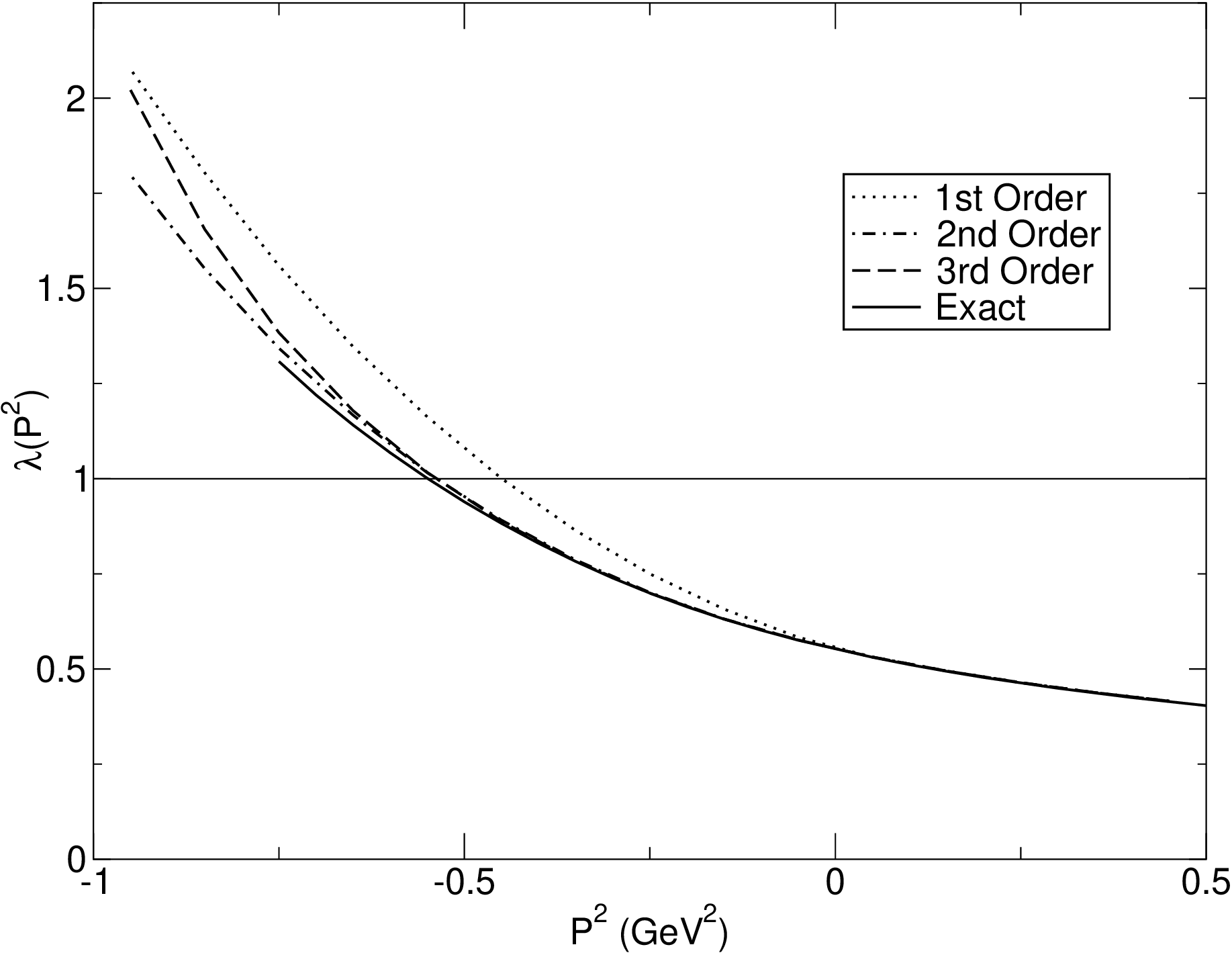}
\caption{The BSE eigenvalue in the $\rho$ channel calculated at the
indicated orders in the Taylor expansion compared to the model exact
result~\protect\cite{Maris:1999nt}.  The physical point is at
$\lambda=1$.
\label{fig:eigenv} }
\end{figure}

For the $\rho$, we show in Fig.~\ref{fig:eigenv} the BSE eigenvalue
$\lambda(P^2)$ for the first three orders of the Taylor expansion.  We
also display the model-exact behavior obtained by direct integration
throughout the relevant complex domain.  At the physical point
\mbox{$\lambda =1$}, a converged value of the mass is evident in the
figure at second order in the Taylor expansion method.  However, the
figure also suggests that at larger timelike $P^2$ a converged
behavior is not yet clearly evident at third order.

The results in Table~\ref{tab:resvecnew} for $\rho, K^\star$ and
$\phi$ illustrate this in more detail.  At second order the masses and
$f_\rho$ are within 1\% of the model exact results while $f_{K^\star}$
and $f_\phi$ are within 5\% and 3\% respectively.  Since $K^\star$ is
not a charge conjugation eigenstate, the solution of the BSE is more
difficult due the the lack of definite parity in the variable $q \cdot
P$.  Furthermore, one expects larger errors in the decay constants
than in the masses since the former require normalization of the BS
amplitudes via Eq.~(\ref{gennorm}) and the derivatives therein add to
the numerical difficulty.  The third order results for the masses
improve slightly for $\rho$ and $K^\star$ but $m_\phi$ deteriorates
slightly to acquire a 2\% error; the decay constants remain within 5\%
of the exact values.  The results do not indicate that the Taylor
expansion has yet converged and this correlates with the behavior
evident in Fig.~\ref{fig:eigenv}.  Table~\ref{tab:resvecnew} also
makes clear that a truncation of the eight Dirac covariants to just
the dominant one ($\gamma_\mu$) will lead to a persistent error of at
least 15-20\%.  In the case of the $ K^\star$ this tends to raise the
mass to the level where we are unable to determine a converged result.

%
\begin{table}
\caption{\label{tab:resvecnew}
The vector meson masses and electroweak decay constants (in GeV).  Results
are shown through several orders in the Taylor expansion treatment of
$q_\pm^2$ in quark propagator amplitudes.  The effect of truncation  
of the 8 covariants to just the canonical (Dirac matrix) 
covariant is indicated.  Comparison is made with results
from use of quark amplitudes along $\Re(q_\pm^2)$ only, and with the  
model-exact results.  }
\begin{ruledtabular}
\begin{tabular}{l|cccccc}
         &\multicolumn{2}{c}{$\rho$} & \multicolumn{2}{c}{$K^\star$} &
                                        \multicolumn{2}{c}{$\phi$}   \\ 
         & $m_\rho$ & $f_\rho$ & $m_{K^\star}$ 
                        & $f_{K^\star}$ & $m_\phi$ & $f_\phi$   \\ \hline
$1^{st}$ order Taylor   &       &       &       &      &      &     \\ 
All 8 amplitudes        &.669   &.222   &.848   &.276  &1.009 &.299 \\
$V_1$ only              &.811   &.255   & -     & -    &1.202 &.331 \\ \hline
$2^{nd}$ order Taylor   &       &       &       &      &      &     \\ 
All 8 amplitudes        &.732   &.209   &.949   &.253  &1.064 &.266 \\
$V_1$ only              &.856   &.214   & -     & -    &1.235 &.271 \\\hline
%
$3^{rd}$ order Taylor   &       &       &       &      &       &    \\ 
All 8 amplitudes        &.735   &.196   &.934   &.253  &1.054 &.253 \\
$V_1$ only              &.844   &.187   & -     & -    &1.182 &.213 \\ \hline
Real axis only          &.711   &.217   &.872   &.241  &1.066 &.288 \\ \hline
Model exact~\protect\cite{Maris:1999nt}  
                        &.742   &.207   &.936   &.241  &1.072  &.259 \\ \hline
Experiment              &.770   &.216   &.892   &.225  &1.020  &.237 \\ 
\end{tabular}
\end{ruledtabular}
\end{table}

The evident difficulty in use of the Taylor expansion approach as the mass
of the state increases is most likely related to the occurrence of
complex conjugate singularities of the employed quark propagator
amplitudes $\sigma_\alpha(q_\pm^2)$.  In rainbow approximation, which
we employ, the occurrence of complex conjugate logarithmic branch
points with timelike values of $\Re(q_\pm^2)$ is well-known for the
fermion propagator in both QED~\cite{Maris:1992cb,Maris:1994ux} and
QCD~\cite{Stainsby:1992hy,Frank:1996uk}.  For the present model, we
find numerically that the non-analytic behavior nearest to the origin
is at \mbox{$q_\pm^2 = -0.207 \pm i 0.331~{\rm GeV}^2$} for the $u/d$
quark, and at \mbox{$q_\pm^2 = -0.376 \pm i 0.602~{\rm GeV}^2$} for
the $s$ quark.  These points are outside the integration domains
needed for the model-exact $\rho$ and $\phi$ states respectively.
However the critical bound state masses for which these singular
points of the propagators would just begin to enter the integration
domain are $1.09$~GeV for a $\bar u u$ state, and $1.47$~GeV for a
$\bar s s$ state.  A Taylor expansion should diverge at these points;
it is probably the precursor of this that is beginning to show for the
$\rho$ in Fig.~\ref{fig:eigenv}.

\subsection{Strong Decays}
Since the $\rho$ and $\phi$ appear as resonance poles in the timelike
behavior of the charge form factors of the $\pi$ and $K$ mesons, the
pole residues involve the coupling constants $g_{\rho \pi \pi}$ and
$g_{\phi K K}$.  If the charge form factors, or more generally the
$\gamma \pi \pi$ and $\gamma K K$ vertex functions, are formulated in
impulse approximation, then the values of $g_{\rho \pi \pi}$ and
$g_{\phi K K}$ extracted from the pole residues will be in impulse
approximation.  The quality of the latter values will depend directly
upon the quality of the strengths of the $\gamma \pi \pi$ and $\gamma
K K$ vertex functions in the relevant timelike region.  It is known
that the impulse approximation for electromagnetic coupling to mesons
conserves the electromagnetic current (and provides the correct
charge) independent of model parameters as long as the meson BS
amplitudes and photon-quark vertex are in ladder approximation and the
quark propagators are in rainbow approximation~\cite{Maris:2000sk}.
We assume that the dynamics which produces the correct strength at the
photon point will also produce a good quality strength at the mass
shell of the vector mesons and hence render the impulse approximation
good for $g_{\rho \pi \pi}$ and $g_{\phi K K}$.  The extension of this
argument to cover the decay \mbox{$K^{\star+} \to K^+ \pi^0$} is
afforded by the study~\cite{Ji:2001pj} of the semileptonic $K(l3)$
decay \mbox{$K^+ \to l\nu_l \pi^0$} where $K^{\star+}$ appears as a
pole in the $W^+$ vertex.

Here the coupling constants for the decays of $\rho, K^\star$ and
$\phi$ to a pair of pseudoscalar mesons are calculated in impulse
approximation according to Eqs.~(\ref{triangle}-\ref{g_rpp}).  The
constraints on the two external Euclidean momenta $P$ and $Q$ needed
to satisfy the three mass-shell conditions entail an analytic
continuation to a complex value for at least one component.  For
example, in the \mbox{$\rho \to \pi \pi$} case, with the momenta
defined in Eq.~(\ref{triangle}), we have \mbox{$Q=(i m_\rho,
\vec{0})$}, \mbox{$P^2 = $} \mbox{$m_\rho^2/4 - m_\pi^2$} and \mbox{$P
\cdot Q = 0$}.  A by-product of this is that two of the quark
propagators and two of the meson BS amplitudes are needed in domains
of the complex $q^2$-plane.  In all cases we employ second order
Taylor expansions about the closest point on the positive real axis.
We also test the dependence upon the number of Dirac covariants
retained for the vector and pseudoscalar BS amplitudes.  With a
maximum of eight for the vector and four each for the pseudoscalar BS
amplitudes, there are a maximum 128 distinct quark loop integrals to
be performed.

The results are summarized in Table~\ref{tab:ccres} in comparison with
experimental values obtained from decay widths.  With just the
dominant covariant employed for each meson ($\gamma_5$ for
pseudoscalars, $\gamma_\mu$ for vectors, i.e, (v,p)= (1,1) in
Table~\ref{tab:ccres}) the coupling constants that can be obtained are
about 50\% larger than experiment.  It is known from work on the
vector meson BSE that a very efficient truncation for soft physics can
be obtained with a particular set of 5 covariants~\cite{Maris:1999nt}.
With use of this truncation together with all 4 covariants for each
pseudoscalar, the results in Table~\ref{tab:ccres} indicate that the
dominant physics has been captured; the $K^\star$ decay is within 1\%,
the $\phi$ is within 12\% and the $\rho$ decay is 18\% less than
experiment.  The addition of the remaining three transverse vector
covariants leads to modest improvement giving a deviation from
experiment of between 5\% and 10\% with the error being larger if the
vector meson is lighter.

\begin{table}
\caption{\label{tab:ccres} Vector decay coupling constants calculated
with the imaginary parts of amplitudes treated through order two in
the Taylor expansion.  The dependence on the number of invariant
amplitudes employed is indicated.  The $K^\star$ decay process shown
in brackets is related by isospin symmetry to the former process by a
factor $1/\sqrt{2}$.  The notation (v,p) indicates the number of
invariant amplitudes used for the vector meson and for the
pseudoscalar mesons respectively.  }
\begin{ruledtabular}
\begin{tabular}{c|c|c|c|c}
$g_{v \to p \, p}$  &(v,p)$=$(8,4)&(v,p)$=$(5,4)& (v,p)$=$(1,1)&Expt \\ \hline
$g_{\rho \to \pi\pi}$      &5.14          & 4.93     &8.8    &6.02     \\ 
$g_{\phi \to KK}$          &4.25          & 4.06     &6.91   &4.64     \\ 
$g_{K^{\star+} \to K^0\pi^+}$  &4.81      & 4.56     & -     &4.60     \\
($g_{K^{\star+} \to K^+\pi^0}$)&(3.40)    & (3.22)   & -     &(3.26)   \\ 
\end{tabular}
\end{ruledtabular}
\end{table}

Since the width of the $\rho$ is almost 20\% of its mass while the
widths of the $\phi$ and $K^\star$ are significantly less important,
we expect the ladder approximation for the BSE kernel (which omits the
strong channels $\pi \pi$, $K K$ and $K \pi$ respectively) to be less
accurate for the $\rho$ than for the $\phi$ and $K^\star$.
Accordingly we speculate that this is largely the reason why the
result for $g_{\rho \to \pi\pi}$ in Table~\ref{tab:ccres} deviates
from experiment twice as much (15\%) as do the other decay constants.

The results shown in Table~\ref{tab:ccres} compare well with coupling
constants extracted from the behavior of electroweak form factors
calculated, independently and in the same model, at timelike momentum
near the vector meson poles~\cite{Maris:2001rq,Maris:2001am}.  The
general behavior of a pseudoscalar meson charge form factor
$F_P(Q^2)$ in this domain is~\cite{Maris:2001rq}
\begin{equation}
 F_P(Q^2) \to \frac{ g_{v\to p p} \; m_v^2}
        { g_v \,  (Q^2 + m_v^2 - i m_v \, \tilde{\Gamma}_v) } \;.
 \label{rhopole}
\end{equation}
Here \mbox{$m_v^2/g_v$} is the \mbox{$v$-$\gamma$} coupling strength
calculated from the \mbox{$v \to e^+ \, e^-$} decay, and
$\tilde{\Gamma}_v$ is the vector meson width associated with the
mass-pole.  In the present work the ladder approximation to the BSE
produces real mass-poles.  The vector decay coupling constants
obtained in this way from a fit to the calculated $F_\pi(Q^2)$ and
$F_K(Q^2)$ form factors and from the form factor for the semileptonic
$K(l3)$ decay \mbox{$K^+ \to l\nu_l \pi^0$} are shown in
Table~\ref{tab:polefit}.  These results from a pole fit employ all
relevant covariants for BS calculations and employ complex plane
continuations where necessary.  From this perspective the comparison
with the left-most column of Table~\ref{tab:ccres} provides a rough
measure of the effectiveness of the Taylor expansion method for BSE
solutions.

\begin{table}
\caption{\label{tab:polefit} The present results for the coupling
constants $g_{v \to p p}$ (calculated with all covariants) compared to
values extracted from a pole fit to timelike electroweak form 
factors, and also to experiment.}
\begin{ruledtabular}
\begin{tabular}{c|ccc}
    $g_{v \to p p}$          &   this work  & pole
fit~\protect\cite{Maris:2001rq,Maris:2001am}         & Expt  \\ \hline
$g_{\rho \to \pi\pi}$        &    5.14      & 5.2    & 6.02   \\
$g_{\phi \to KK}$            &    4.25      & 4.3    & 4.64   \\
$g_{K^{\star+}\to K^0 \pi^+}$ &   4.81      & 4.1    & 4.60 
%
\end{tabular}
\end{ruledtabular}
\end{table}
%

\section{Estimate of Heavier States}
In the present formulation, the nearby singularities discussed in
Sec.~\ref{masses} hinder a direct application of our rainbow-ladder
model for masses above about 1.2~GeV.  In particular, it is of
interest to go beyond the present S-wave states to consider orbital
excitations such as the axial vectors $a_1$ and $b_1$.  One can
however provide an indirect estimate as follows.  Although the 
quark propagator amplitudes $\sigma_\alpha(q_\pm^2)$ are
complex-valued in the complex $q_\pm^2$ plane, the ladder BSE
eigenvalue $\lambda(P^2)$ is real and so is the mass.  The evident
cancellations may allow the real axis behavior of the quark amplitudes
to dominate the outcome.  As a test, we introduce a real-axis
approximation by setting the imaginary part of the momentum argument
of the quark propagator amplitudes $\sigma_\alpha(q_\pm^2)$ to zero in
the BSE kernel.  Then, instead of Eq.~(\ref{taylor}), one has
\begin{eqnarray}
\label{realaxis}
 \sigma_\alpha(q_\pm^2)  &\approx& \sigma_\alpha(q^2 - m^2/4)  \; ,      
\end{eqnarray}
where \mbox{$\alpha=v,s$} and \mbox{$q^2 > 0$}.  This is not quite
identical to zeroth order in the Taylor expansion of
Eq.~(\ref{taylor}) because the latter expands about the nearest point
on the positive real axis, while in Eq.~(\ref{realaxis}) the interval
$[- m^2/4, \infty]$ is used.  Thus the required domain of the timelike
real axis is treated exactly and this can be important for the heavier
mesons.

To implement this approximation, we use the quark DSE to continue the
propagator solutions to the timelike (negative real) $p^2$ axis.  In this
respect, it is advantageous to enhance numerical stability of the
present model by modifying the infrared behavior of the factor ${\cal
F}(k^2)$ in the small second term of the effective coupling, given in
Eq.~(\ref{gvk2}), so that, like the first term, it is identically zero
at \mbox{$k^2=0$}.  Its UV limit and range should remain much the
same.  To achieve this we use
\begin{equation}
\tilde{{\cal F}}(s)=\frac{1 - {\rm e}^{-(s/4 m_t^2)^2}  } {s}~~.
\label{mod_F}
\end{equation}
With the real axis approximation of Eq.~(\ref{realaxis}), the
resulting modified model yields mass estimates that tend to become more
accurate as the mass increases.  From Table~\ref{tab:PSresults}
$m_{\pi/K}$ are underestimated by 10\% and $f_{\pi/K}$ are
underestimated by 20\% compared to the model-exact values.  However,
as shown in Table~\ref{tab:resvecnew}, the produced ground state
vector meson masses $m_\rho$, $m_{K^\star}$ and $m_\phi$ are within
4\%, 7\% and 1\% of the model-exact values respectively; the decay
constants deviate by 5\%, 0\% and 11\% respectively.  We therefore
expect this real-axis approximation to provide mass estimates in the
1-2 GeV range with an error of about 5-10\%.
 
For a $u/d$ quark axial vector state, the required general form of BS
amplitude is simply $\gamma_5$ times Eq.~(\ref{vecBSAform}).  For the
$a_1$ ($1^{++}$), amplitudes $V_5$, $V_7$ and $V_8$ will be odd in $q
\cdot P$ and the rest will be even; for the opposite $C$-parity state
$b_1$ ($1^{+-}$) the opposite is true.  The real axis approximation
gives the following axial vector estimates: \mbox{$m_{a_1} \approx$}
0.891~GeV and \mbox{$m_{b_1} \approx$} 0.775~GeV.  Hence the orbital
excitation energy \mbox{$m_{a_1} - m_\rho$} is 0.150~GeV and thus a
factor of 3 too small in this model.  Indications from estimates
attempted using the complex plane information are similar, as are
recent results from a related study~\cite{Alkofer:2002bp}.  Models of
the present rainbow-ladder type are significantly too attractive for
these orbital excitations.  Other studies of $a_1$ and $b_1$ based on
the DSEs have used a separable approximation where the quark
propagators are the phenomenological
instruments~\cite{Burden:1997fq,Bloch:1999vk,PichExotic02} and these
studies find more acceptable masses in the vicinity of 1.3~GeV.  The
manner in which the effective interaction is modeled is clearly
important.  Preliminary results from an extension of the present
rainbow-ladder level through a 1-loop dressing of the quark-gluon
vertex, while preserving the vector and axial vector WTIs, indicate
that the separation of the axial vector states from the vector states
increases significantly and becomes quite acceptable~\cite{Watson_PCT02}.

Recently there has been interest in meson states with quantum numbers
that are called exotic in the sense that they cannot be produced as
$\bar q q$ bound states within static quantum mechanics where
$C$-parity is given by \mbox{$C= (-1)^{L+S}$}.  The $\pi_1$(1400)
state, formerly called the $\hat{\rho}(1405)$, is a resonance with
\mbox{$J^{PC}=1^{-+}$} that is supported by evidence from $\pi d$
scattering and $\bar p d$ annihilation~\cite{Groom:2000in}.  In a
covariant field theory treatment of bound states via the BSE, the
extra degree of freedom represented by relative time, or relative
energy, allows states to have an additional classification in terms
of an associated time-parity quantum number $\kappa$ and $C$-parity 
is given by the more flexible form \mbox{$C= (-1)^{L+S+\kappa}$}.  
The states with \mbox{$\kappa={\rm odd}$} have no static quantum 
mechanical $\bar q q$ analog and
in this sense the exotic $J^{PC}$ obtainable from the ladder BSE are
evidently a realization of the multi-particle features of covariance.
Although a full understanding within field theory of the norm of these
odd solutions of the BSE remains to be achieved~\cite{PichExotic02},
examples from otherwise realistic models are of interest in connection
within on-going experimental searches.

A covariant separable BSE model, closely connected with the present
model, has recently produced~\cite{PichExotic02} \mbox{$m_{\pi_1} =
1.439$}~GeV, i.e., about 100~MeV above the $a_1$, which in
that model is \mbox{$m_{a_1} =$} \mbox{$ 1.337$}~GeV.  To estimate 
whether this is consistent with the
present study, we use the real axis approximation to search for a
\mbox{$J^{PC}=1^{-+}$} solution.  The general form for such a BS
amplitude is the same as Eq.~(\ref{vecBSAform}) for the $1^{--}$
amplitude except that one reverses the odd-even property of the
invariant amplitudes in $q \cdot P$.  That is, $V_3$ and $V_6$ are to
be even, the rest are to be odd.  This produces a $\pi_1$ solution
112~MeV above the $a_1$ within the same approach.  In this respect we
agree with Ref.~\cite{PichExotic02}.  We speculate that if the problem
of the ground state orbital excitations being about 300~MeV too low is
remedied, then the $\pi_1$ state in this model should rise by a
similar amount to be around 1.3-1.4~GeV.

\section{\label{sec:discussion} Discussion}
We have studied vector meson strong decays within a Euclidean space
model of QCD based on the Dyson--Schwinger equations truncated to
ladder-rainbow level.  The infrared structure of the ladder-rainbow
kernel is described by two parameters; the ultraviolet behavior is
fixed by the one-loop renormalization group behavior of QCD.  Within
the $u/d$ and $s$ quark sector we have obtained the coupling constants
for the strong decays: \mbox{$\rho \to \pi \pi$}, \mbox{$K^\star \to K
\pi $}, and \mbox{$\phi \to K K$}.  The deviation from experiment is
between 5\% and 10\% with the error being larger if the vector meson
is lighter.
  
Our method of calculation employed a Taylor expansion to continue the
dressed quark propagators from the real spacelike $p^2$ axis into the
domain of the complex plane required by the mass-shell condition of
the meson bound states.  We used this both for the solution of the BSE
and for calculation of decays in the impulse approximation.  The
benefit is that knowledge of the dressed quark propagators is needed
only along the real spacelike $p^2$ axis and this simplifies the
numerical requirements considerably.  The vector and pseudoscalar
meson masses compared well with the model-exact results obtained
previously by direct continuation of the dressed quark propagators
into the complex plane via the DSE.  In particular, the masses and
electroweak decay constants at second order are all within 5\% of the
model-exact results.  The increased deviation observed at third order
and for heavier masses is attributed to nearby complex conjugate
singularities in the propagators that are known to occur in rainbow
solutions of the DSE.

For mesons heavier than the $\phi$, these singularities hinder the
applicability of Euclidean space implementations of the DSEs as
presently formulated.  The Taylor expansion method cannot lessen this
difficulty.  We therefore explored a real-axis approximation that
exploited the cancellations of imaginary parts evident in the way the
complex-valued quark amplitudes off the real axis enter the BSE kernel
to produce real eigenvalues and masses.  The real-axis approximation 
avoids complex plane singularities and, based on its performance for
the pseudoscalar and vector mesons, we conclude it can be used for mass
estimates in the 1-2~GeV region with an accuracy of about 5-10\%. 
We then produced
estimates of $m_{a_1}$ and $m_{b_1}$ for orbital excitations as well
as $m_{\pi_1}$ for the proposed exotic $1^{-+}$ vector state.  We
mentioned evidence that an extension beyond ladder-rainbow level to
include dressing of the quark-gluon vertex could address our finding
that the present model is too attractive for states just above the
ground state vectors.  With this, our estimate for $m_{\pi_1}$ is
1.3-1.4~GeV.

Work in progress~\cite{PichTand02} on the case of propagators having
explicit complex conjugate poles, indicates that a reformulation
produces well-defined integrals and integral equations above the
critical masses.  Beyond rainbow-ladder approximation, very little is
known about the singularity structure.  In one simplified model that
includes a fully dressed quark-gluon vertex in a way that has allowed
exploration of analytic properties, the non-analytic behavior is
relegated to essential singularities at infinity~\cite{Burden:1992gd}.


\begin{acknowledgments}
This work was funded by the National Science Foundation under Grant
Nos.\ PHY-9722429 and PHY-0071361; and by the Department of Energy under 
Grant Nos.\ DE-FG02-96ER40947 and DE-FG02-97ER41048.  We acknowledge a 
grant of resources from  the Ohio Supercomputer Center, Columbus, Ohio.   
The authors are grateful to Patrick Bowman and Tony Williams for kindly 
providing the lattice-QCD data and useful advice on usage.

\end{acknowledgments}

\bibliography{refsPM,refsPCT,refsCDR,refs}

\end{document}